\documentclass[
superscriptaddress,
pra,
showpacs,
showkeys,
aps,
floatfix,
]{revtex4-1}

\usepackage{amssymb}
\usepackage{graphicx}
\usepackage{bm,bbm}
\usepackage{color}

\newcommand{\eg}{\textit{e.g.}}

\newcommand{\rmw}{\mathrm{w}}

\providecommand{\One}{\leavevmode\hbox{\small1\kern-3.8pt\normalsize1}}
\newcommand{\ie}{\textit{i.e.} }
\newcommand{\mt}{\mathrm }

\newcommand{\omo}{\omega_{\mathrm{o}}}

\newcommand{\rme}{\mathrm{e}}
\newcommand{\rmi}{\mathrm{i}}
\newcommand{\rmd}{\mathrm{d}}

\newcommand{\h}{\hat}

\newcommand{\la}{\langle}
\newcommand{\ra}{\rangle}

\newcommand{\bA}{\bm{A}} 
\newcommand{\bM}{\bm{M}} 
\newcommand{\bB}{\bm{B}} 
\newcommand{\bN}{\bm{N}} 
\newcommand{\bR}{\bm{R}} 
\newcommand{\bC}{\bm{C}}

\newcommand{\gammo}{\gamma_{\mathrm{o}}}

\begin{document}

\title[Analytic solutions to various dissipation models of the simple and driven quantum harmonic oscillator]
{Analytic solutions to various dissipation models of the simple and driven quantum harmonic oscillator}

\author{P. C. L\' opez V\' azquez}
\email{pablo.lopez@valles.udg.mx}
\author{R. Santos-Silva}
\email{roberto.santos@academicos.udg.mx}
\affiliation{Departamento de Ciencias Naturales y Exactas, Universidad de Guadalajara,
Carretera Guadalajara-Ameca Km 45.5, C.P. 46600, Ameca, Jalisco, M\'exico}

\begin{abstract}
We obtain analytic solutions to various models of dissipation 
of the quantum harmonic oscillator, employing a simple method
in the Wigner's function Fourier transform description 
of the system; and study as an exemplification, 
the driven open quantum harmonic oscillator. 
The environmental models we use are based on optical 
master equations for the zero and finite temperature bath and 
whose open dynamics are described by a Lindblad master equation,
and also we use the Caldeira-Leggett model for the high temperature limit,
in the the under damped an the over damped case.
Under the Wigner's Fourier transform
or chord function as it has been called,
it becomes particularly simple to solve 
the dynamics of the open oscillator in the sense that the dynamics 
of the system are reduced to the application of an evolution matrix
related to the damped motion of the oscillator.
\end{abstract}
\maketitle


\section{\label{int}Introduction}
Open dynamics in quantum systems have become a fundamental 
part of the quantum theory due to the inherent interaction 
with the surroundings and thus, any theory that boasts of 
being a truthful approximation of reality, should include 
dissipation. The shape of the world itself may have its 
foundations on dissipation and complexity, for instance,
decoherence and the emergence of classicallity are 
examples that the quantum world may be inherently 
open~\cite{Zu81,Zu82,Jo85,Pa93,Ga96}.
On the other hand, periodic motion is
widely present in nature and for that reason, 
the harmonic oscillator has been one most employed
systems used to study the environmental effects on a quantum
system~\cite{De81,Sa87,Is94,De16}. The open harmonic oscillator 
is deeply involved in many fields of the quantum theory,
\eg~quantum and atom optics, molecular dynamics etc.; 
furthermore, the development of quantum technologies such as quantum
computation~\cite{NiCh10}, super conducting 
quantum devices~\cite{Jo06,Gi09}, nano electro-mechanical 
systems~\cite{Go98,Do04,Ut06} and opto-mechanical
quantum systems~\cite{Ve12,As14,Sa17} are also fundamentally based on the dynamics 
of the open harmonic oscillator.
The treatment to tackle open dynamics 
has been standardized with the formulation of interacting systems,
in which a central system, which in our case 
is the harmonic oscillator, interacts with an idealized 
reservoir \eg~a very large collection of harmonic oscillators
at a certain temperature, or at a zero temperature,
the later corresponding to a vacuum field 
which will have the effect 
to only produce damping to the oscillator. 
In many situations, the environment is assumed to be stationary 
and thermalized, in such a way that any possible perturbation produced 
from the central system, thermalizes at time scales much more shorter that
the characteristic times of the central system dynamics and  
hence the conception of the reservoir. Other models
of dissipation conceptualize the environment as not being 
a large  thermalized system within the idea of 
what a reservoir should be, but rather they are considered 
as a small subsystems which could exchange processes with the central 
system. In this case the dynamics of the central system 
are considered Non-Markovian~\cite{Ri14,La10}, and its said that 
certain amount of information that is lost in the environment,
could return to the central system after some finite time. 
One of the most popular ways to treat the open dynamics 
of the system is by employing master equations
which have a specific form, depending on the approximations 
and assumptions done to the interaction between 
the central system and the environment and also 
on the characteristics of the environment such 
as the temperature, the strength of coupling, 
the spectral density of the oscillators that constitute 
the environment, and also on approximations done at the derivation of the 
master equations such as the weak coupling limit and the
Born-Markov approximations~\cite{Br02}.
In this paper we will provide analytic solutions to a class 
of Markovian systems: the zero-finite temperature model
based on optical interactions for which 
the master equation posses a Lindblad form~\cite{Br02}, 
and the high temperature model, also known as
the Caldeira-Leggett model~\cite{Ca83}. For the later we will 
focus on the under damped  and the 
over damped regime~\cite{An03}. 
We will also give analytic solutions of the driven 
harmonic oscillator when the driving is done 
due to an external periodic force applied to the
oscillator~\cite{Pi06}.
For deriving the analytical solutions, 
we will make use of the Wigner function Fourier's transform 
or chord function as it has been called~\cite{Ozo98,Ozo02,Pr17}. Within 
this approximation the solution of the different models becomes 
a standard method as long as the central system is an integrable 
system and the solutions are based on the application of an evolution 
matrix of the classical damped harmonic oscillator. The main 
propose of this paper is to provide with a standard method 
of solution, the dynamics of the open harmonic oscillator 
or any system whose classical counterpart posses dynamics 
that can be described with the help of an evolution matrix.
The paper is organized as follows: In section \ref{me}, we give 
a briefly review of the open harmonic oscillator and
define the chord function representation of the master
equation. In section \ref{zft} we discuss about the 
finite temperature model and give an analytic solutions
to it and to its different limits. In section \ref{CL}
we tackle the high temperature cases based on
the Caldeira-Leggett model for the under damped and 
over damped regimes, in section \ref{df} we give analytical 
solutions of the driven open oscillator when the environment
given by the finite temperature and the high temperature models.
Finally in section \ref{sum} we give
a summary of our results.

\section{\label{me} Master equations}
The dynamics of an open system are 
generally described through a master equation 
of the reduced system which consist 
on two parts; one concerning the von Neumann dynamics where
the unitary dynamics are contained,
and the dissipative dynamics due to the interaction with
the environment.  In a dimensionless description frame,
a master equation for the harmonic oscillator 
will typically have the following form:
\begin{equation}\label{meg}
\rmi\, \frac{\rmd\varrho}{\rmd \tau} =
  [H_{\mt{osc}},\varrho] + \rmi \mathcal{A}[\varrho]
\end{equation}
where $H_{\mt {osc}}=\left(\h{x}^2 + \h{p}^2 \right)/2$
is the Hamiltonian of the quantum harmonic oscillator,
in the dimensionless description:
$\h{X} = \sqrt{\hbar/(m \omo)}\, \h{x}$ and 
$\h{P} = \sqrt{\hbar m \omo}\, \h{p}$, where 
$\h{X}$ and $\h{P}$ are the position and momentum operators
and $\omo$ is the angular frequency of the oscillator, and
we measure everything in a dimensionless
time scale $\tau = \omo t$.
The superoperator $\mathcal{A}[\cdot]$ describing
the dissipative part of the dynamics may be chosen accordingly 
to the model of dissipation we will be interested to show.
We will focus on two different models of dissipation which 
have been vastly used in the literature: the high temperature model,
implemented with the Caldeira-Leggett model~\cite{Ca83}, and the finite 
or zero temperature models
which are based described through quantum optical 
master equations~\cite{Br02}.
Within the literature, one can find different ways 
to solve or approximately solve the master equations,
one particular and simple way to do it is by employing the 
the Fourier transform of the Wigner function also called 
chord function~\cite{Ozo98,Ozo02},
or characteristic function in quantum optics~\cite{Ge04}.
In the following we 
give a brief description of the chord function and
some of the advantages of working with it.

\subsection{The chord function representation}
The Fourier transform of the Wigner function
or the chord function~\cite{Ozo98,Ozo02} is defined as follows:
\begin{eqnarray}\label{wtr}
\rmw(k,s; \tau)&=& \int_{-\infty}^{\infty}\int_{-\infty}^{\infty}\rmd p\, \rmd q
  \; \rme^{\rmi k q +\rmi s p}\, \mt{W}(q,p; \tau)\\
&=& \int_{-\infty}^{\infty}\rmd q\, \rme^{\rmi k q}\,
      \la q+s/2 |\,  \varrho(\tau)\,  | q-s/2\ra \; .
\end{eqnarray}
Within this description, the
the evolution of observables can be obtained by doing: 
\begin{equation}
  \langle \hat{A}(\tau) \rangle  = \int_{-\infty}^{\infty}\int_{-\infty}^{\infty}\, \rmd k\, \rmd s\, 
 \rmw(k,s,\tau)\, \bar{A}(k,s)
\end{equation}
where $\bar{A}(k,s)$ is the Fourier transform of the Weyl
symbol of operator $\hat{A}$~\cite{weyl27,Kl09}. 
Particularly, the evolution of
the energy of the oscillator in terms of the chord function, 
one has the following relation:
\begin{equation}\label{aven}
  E(\tau)  =   \langle H (\tau) \rangle =  
  -{1 \over 2}\left( \partial^2_k + \partial^2_s\right) \rmw(k,s,\tau)\bigg|_{k=0, s=0}\,.
\end{equation}
Within the chord function description, the position and momentum probabilities 
distributions can be easily obtained: 
$P(q,\tau) =  \int_{-\infty}^{\infty} \rmd k \, \rmw (k,0)\,\rme^{-\rmi k q}$,
and $P(p,\tau) = \int_{-\infty}^{\infty} \rmd s \, \rmw (0,s)\,\rme^{-\rmi s p}$; but 
probably the most useful feature about the chord function description
is the simplicity in solving the dynamics of the open oscillator 
since the second order partial differential equations 
appearing in the Fokker-Planck equations, 
are transformed to first order partial differential equations
as long as the central system is integrable.
In this description, the operator dependent master equation
is mapped into the chord space $\{k,s\}$ where the dynamics 
of the open oscillator are obtained by the application of an evolution matrix
related to the damped motion of the oscillator.

\section{Solutions to the dissipative models}
\subsection{\label{zft} Finite temperature}
The finite temperature environmental models
are based on quantum optical interactions such as photon exchange processes 
for which the interaction Hamiltonian is 
formulated in terms of Jaynes-Cummings model~\cite{Kl09}:
$ H_{\mt{FT}} = \sum_i g_i \h{a}^{\dag} \h{b}_i + g^{*}_i \h{a} \h{b}_i^{\dag}$, 
where $\h{a}, \h{a}^{\dag}$ and $\h{b}_k, \h{b}_k^{\dag}$  are the
annihilation and creation operators of the central harmonic oscillator
and the $kth$ oscillator of the bath respectively, while the $g_i$'s
are the coupling strengths of the central oscillator to the $i$-th oscillator in the bath.
A master equation in the optical quantum limit can be derived 
under standard microscopic derivations~\cite{Br02}, such as the weak coupling which
consist on taking the interaction between the system
and the environment in a perturbative regime, and the Born-Markov approximation
which assumes a Markovian type of environment with a flat spectral density 
and an upper cut off frequency which determines the range of influence 
of the environment to the system. The resulting master equation has
a Lindblad form:
$\rmi\, \rmd\varrho/\rmd \tau =
  [ H_{\mt{osc}},\varrho] + \rmi \mathcal{L}[\varrho]$, 
where $\mathcal{L}[\cdot]$ is a super operator correspondent to the non-unitary evolution 
and hence the dissipative processes in the system. In this particular case 
it has the following form:
\begin{eqnarray}\nonumber
\mathcal{L}[\varrho] &=& - \gamma \left(1+\bar{n}\right)\left(a^{\dag}a\, \varrho 
- 2\,a\varrho\,a^{\dag} + \varrho\, a^{\dag} a\right)
- \gamma\, \bar{n}\left(aa^{\dag}\, \varrho - 2\,a^{\dag}\varrho\,a + 
\varrho\, a a^{\dag}\right)
\end{eqnarray}
where $a^{\dag}(a)$ are the creation(annihilation) operators and $\bar{n}$ is the 
Planck distribution function $\bar{n} = \left(\rme ^{1/D} - 1\right)^{-1}$,
and  $D = {k_{\rm B}\, T / (\hbar\, \omo)}$ is the 
dimensionless diffusion constant and $\gamma= \gammo/\omo$ 
the dimensionless coupling rate to the environment.
In terms of the chord function description, the master
equation has the following form:
\begin{equation}\label{mechft}
 \partial_{\tau}\rmw  + (s + \gamma k )\partial_k \rmw - (k-\gamma s)\partial_s\rmw 
 = - {\gamma_{+}\over 2} (k^2 + s^2) \rmw \\ 
\end{equation}
where $\rmw=\rmw (\vec{r})=\rmw(k,s,\tau)$, and we have defined 
$\gamma_{+} = 2 \gamma (\bar{n} + 1/2)$. 
Therefore, the master equation has become a first order partial differential
that  can be solved exactly with standards methods.
We use the method of the characteristics in which 
one can put the PDE into a set of ordinary
parametric differential equations:
\begin{eqnarray} \label{paraft1}
 {\rmd k \over \rmd\tau}  & = & s + \gamma k,\\ \label{paraft2}
 {\rmd s \over \rmd\tau}  & = & -k + \gamma s,\\  \label{paraft3}
 {\rmd \rmw \over \rmd \tau} & = & -{\gamma\over 2} (k^2 + s^2)\rmw  \,.
\end{eqnarray}
The first pair of equations can be placed together into a second 
order ordinary differential equation: $\ddot{k} -2\gamma\dot{k} + \kappa^2 k = 0 $,
where  $\kappa^2 = 1 + \gamma^2$ which unlike to the high temperature models;
there is no mathematical restriction on the values that 
$\gamma$ can take (see section \ref{CL}). 
The solution of the second order differential equation
can be written in general terms as:
 \begin{equation}\label{sk}
k(\tau) = \rme^{\gamma\tau/}\, \big ( a_1\, \sin\tau + a_2\,
               \cos\tau\big )
\end{equation}
where $a_1$ and $a_2$ are the characteristic curves which remain constant for all time.
The variable $s(\tau)$ may be obtained through the equation (\ref{paraft1}),
$s = \dot{k} - \gamma\, k$ yielding:
\begin{equation}
 s(\tau) = \rme^{\gamma\tau/2}\, \big (  a_1\, \cos\tau - a_2\,\sin\tau\big )\,. 
\end{equation}
Once the time dependence of the variables is solved, one can easily construct 
an evolution matrix $\bR$ which will describe the map along the
characteristics of any point described by the vector $\vec{r}(\tau) = (k(\tau),s(\tau))$,
at the time $\tau$, to any other point 
$\vec{r}(\tau + \sigma) = (k(\tau + \sigma),s(\tau+\sigma))$
at time $\tau + \sigma$, \ie\,$\vec{r}(\tau+\sigma) = \bR(\sigma)\,\vec{r}(\tau)$, 
where
\begin{equation}\label{Matft}
\bR(\sigma) = \rme^{\gamma\sigma}\left(\begin{array}{cc}
   \cos \sigma  & \sin \sigma \\
   -\sin \sigma & \cos \sigma 
   \end{array}\right)\;.
\end{equation}
The map $\bR$ is invertible \ie~$\bR^{-1}(\tau) = \bR(-\tau)$ and posses 
group properties: $\bR(\tau)\bR(\tau') = \bR(\tau + \tau')$.
Also, notice the independence of matrix $\bR$ on the angular frequency $\kappa$.
This indirectly implies that the under damped and the over damped 
regimes may all be contained in this map. The time 
integration over the last equation of (\ref{paraft3}) will be in the form 
\begin{equation}\label{intft}
\int_{\rmw (\tau)}^{\rmw (\tau+\sigma)}{\rmd\rmw \over \rmw }
=-{\gamma_{+}\over 2}\int_{\tau}^{\tau+\sigma}\rmd\tau ' \,   \left( k^2(\tau ') +  s^2(\tau ')\right)\; .
\end{equation}
where $\rmw(\tau + \sigma)$ and $\rmw(\tau)$ are the chord functions at time $\tau + \sigma$ and $\tau$
respectively. Within this map, the second term of the r.h.s can be calculated by using the fact that
\begin{eqnarray}
 k(\tau')  &= & R_{11}(\tau'-\tau)k(\tau) + R_{12}(\tau'-\tau)s(\tau)\\
 s(\tau')  &=& R_{21}(\tau'-\tau)k(\tau) + R_{22}(\tau'-\tau)s(\tau)\,,
\end{eqnarray}
where $R_{ij}$ are matrix elements of the map described in (\ref{Matft}),
thus one can write down an explicitly expression for the 
evolution of this chord function matrix element from some initial time $\tau$ to an arbitrary
final time $\tau+\sigma$:
\begin{equation}\label{solwft}
\rmw(\vec{r},\tau+\sigma)= 
\rmw\big(\, \bR(-\sigma) \vec{r}\, ,\, \tau \big)\exp\left( - {\gamma_{+}\over 2}\,\alpha(\sigma)\, |\vec{r}\,|^2\, \right)
\end{equation}
where $\alpha(\sigma)$ is given by 
\begin{equation}\label{tfft}
 \alpha(\sigma) = \int_{0}^{\sigma} \rmd \tau'\left(R^{'2}_{11}(-\tau') + R^{'2}_{12}(-\tau') \right)
 ={1-\rme^{-2\gamma \sigma}\over 2\gamma}\, . 
\end{equation}
The solution (\ref{solwft}) represents 
the evolution of the  chord function from some initial time $\tau$ to an arbitrary
final time $\tau+\sigma$. 
Probably a more familiar way to write down the solution is
in the Wigner's function representation of the density matrix. For doing so, 
one would need to specify the initial conditions of the oscillator. 
Assuming a general coherent initial state:
$\psi (x)  =  \exp\left(\rmi(x-x_o) - (x-x_o)^2/4 \right)/(2\pi)^{1/4}$
with $x_o$ and $p_o$ being the initial averaged position and momentum of the oscillator, then  
by doing the inverse transformation  of (\ref{wtr}), the Wigner function reads:
\begin{eqnarray}\label{Wft}
 \mathrm{W}(q,p,\sigma) &=& {1 \over 2\xi(\sigma)}
 \exp \left[ -{\left( q - x_o(\sigma)\right)^2\over 4\xi(\sigma)} -{\left( p - p_o(\sigma)\right)^2\over 4\xi(\sigma) } \right]
\end{eqnarray}
where $x_o(\sigma)$ and $p_o(\sigma)$ are the components of the vector $\vec{r}_o=(x_o,p_o)$
that has been mapped with the transpose of the matrix $\bR(-\sigma)$; 
this is $\vec{r}_{o}(\sigma) = \bR^{T}(-\sigma)\vec{r}_o$, and 
$\xi(\sigma) = \rme^{-2\gamma\sigma}/4 + \gamma_{+}\alpha(\sigma)/2$.
The time evolution of the energy  can be obtained by using
equation (\ref{aven}), yielding:
\begin{equation}\label{eft}
 E (\sigma) = E_{\mt o}\, \rme^{-2\gamma\sigma} + \left(\bar{n} + 1/2\right)\left(1- \rme^{-2\gamma \sigma}\right)\,, 
 \end{equation}
where $E_{\mt o}$ is the energy of the initial condition of the oscillator.
From the solution of the energy one can easily sees that as the system
evolves in time, its average energy becomes the dimensionless 
temperature of the environment plus the ground state energy.
This thermalization process can also be observed in (\ref{solwft}), since
for the long time limit; $\lim_{\sigma\rightarrow \infty} \bR(-\sigma)\rightarrow 0$
which implies an independence on the initial condition and at this same limit 
$\alpha(\sigma)\rightarrow 1/2\gamma $ yielding a stationary state in the form
$\rmw_{\mt{st}}(\vec{r})=\lim_{\sigma\rightarrow \infty}\rmw(\vec{r},\tau+\sigma)  
= \exp\left[  - \left(\bar{n} + {1\over 2}\right)\left( k^2 + s^2\right)/2\right]$. 
\subsubsection{\label{ztft}The zero temperature limit}
The zero temperature limit can be taken whenever the 
relation $k_B T << \hbar \omega_o$ is fulfilled.  In such case,
the Planck distribution function $\bar{n}$ appearing inside $\gamma_{+}$, becomes effectively null, 
$\bar{n}(D\rightarrow 0) \rightarrow 0$.
This limit is in correspondence to the damped motion of the oscillator 
alone, and it can be directly taken in the solution for the chord function; in fact the only 
thing to do is the replacement:  $\gamma_{+}\rightarrow \gamma$ yielding:
\begin{equation}\label{solwzt}
\rmw(\vec{r},\tau+\sigma)= 
\rmw\big(\, \bR(-\sigma) \vec{r}\, ,\, \tau \big)
\exp\left( - {\gamma\over 2}\,\alpha(\sigma)\, |\vec{r}\,|^2\, \right)\,,
\end{equation}
while the energy evolves according to:
\begin{equation}
 E (\sigma) = E_{\mt o}\, \rme^{-2\gamma\sigma}  + {1\over 2}\left(1- \rme^{-2\gamma \sigma}\right)\,, 
\end{equation}
and the stationary state becomes the ground 
state:
$\rmw_{\mt{st}}(\vec{r}) = \lim_{\sigma\rightarrow \infty}\rmw(\vec{r},\tau+\sigma)  
= \exp\left[ - {1\over 4} \left( k^2 + s^2 \right)\right]$.
\subsubsection{\label{htft}High temperature approximation}
In the high temperature approximation, one assumes that $D$
is large enough such that one can take the following limit: 
$\lim_{D\rightarrow \infty} ({\bar{n}(D) + 1/2}) \rightarrow D$.
In this case the chord function and the energy becomes:
\begin{equation}\label{solwftht}
\rmw(\vec{r},\tau+\sigma)= 
\rmw\big(\, \bR(-\sigma) \vec{r}\, ,\, \tau \big)
\exp\left( - \gamma \,D\,\alpha(\sigma)\, |\vec{r}\,|^2\, \right)
\end{equation}
and for the energy one has:
\begin{equation}
 E (\sigma) = E_{\mt o}\, \rme^{-2\gamma\sigma} + D\, \left(1- \rme^{-2\gamma \sigma}\right)\,,
\end{equation}
while the stationary state becomes: 
$\rmw_{\mt{st}}(\vec{r}) = \lim_{\sigma\rightarrow \infty}\rmw(\vec{r},\tau+\sigma)  =
\exp\left[-{D\over 2} \left( k^2   +  s^2\right)\right]$.

\subsection{\label{CL} The Caldeira-Leggett model}
The Caldeira-Leggett model consist on a coupling to a
thermal bath which is at high temperatures.
It was first derived as a approximation to the quantization of
the quantum Brownian motion and
it has become one of the most used models for describing 
interaction of a quantum system with a high temperature 
environment.  In this model, the interaction 
Hamiltonian happens through the 
positions of system and the environment:  
$H_{\mt{CL}} = \h{x}\sum_iC_i \h{q}_i$, 
where $\h{q}_i = {1\over \sqrt{2}}( \h{a}^{\dag}_i +\h{a}_i)$
are the position operators
of the oscillators in the the bath and $C_i$ characterizes the
coupling strengths of each of the oscillators to the central system.
It is  worth to mention that the interaction Hamiltonian 
for the finite temperature model, $H_{\mt{FT}}$, can be obtained from
$H_{\mt{CL}}$ by performing a wave rotation
approximation and neglecting the anti rotating terms.
A derivation of a master equation for this type of model  
was first carried out by Leggett et al.~\cite{Ca83}, 
using the path integral formalism of influence integrals,
obtaining a master equation in a a non-Lindblad form:$
\rmi\, \rmd\varrho/\rmd \tau =
  [H_{\mt{osc}},\varrho] + \rmi \mathcal{D}[\varrho]$,
where the superoperator concerning to the non-unitary 
dynamics of the system $\mathcal{D}[\cdot]$, and hence the dissipative 
processes is given by:
\begin{equation}\label{CLo}
 \mathcal{D}[\varrho] =   \frac{\beta}{2}\;
   [\h{x}, \{\h{p},\varrho\} ] - \rmi\beta\, D\, [\h{x}, [\h{x},\varrho]] \; ,
\end{equation}
where $D = {k_{\rm B}\, T / (\hbar\, \omo)}$ is the 
dimensionless diffusion constant and $\beta= 2\gammo/\omo$
is the dimensionless coupling rate to the environment which 
is related to that of the finite temperature 
model as: $\beta = 2\gamma$. 
In the chord function representation 
the master equation becomes first order partial differential equation:
\begin{equation}\label{clchord}
\partial_{\tau}\rmw+(\beta s-k)\partial_s\rmw
+s\partial_k\rmw=-D\beta s^2\rmw\,,
\end{equation}
which also can be cast in a 
parametric form as: 
\begin{eqnarray}\label{ecparcl1}
 {\rmd k\over \rmd \tau}&=&s,\\\label{ecparcl2}
 {\rmd s\over \rmd \tau}&=&\beta s-k,\\\label{ecparcl3}
 {\rmd \rmw \over \rmd \tau}& =& - D \beta s^2\,\rmw\,,
\end{eqnarray}
and by coupling the first two equations, one can write a second order
ordinary differential equation for $k$:
\begin{equation}\label{2ndokcl}
 \ddot{k} -\beta \dot{k} + k = 0\, 
\end{equation}
whose solution can be written in a general form as:
\begin{equation}\label{eck1}
k(\tau)=\rme^{\beta \tau \over 2}\left(  a_1 \sin\omega \tau\, + a_2\cos\omega \tau\right),
\end{equation}
where  $\omega = \sqrt{ 1-\beta^2/4 }$ and 
$a_1$ and $a_2$ are the characteristic curves,
and the solution for $s$ may be obtained from the second equation $s = \dot{k}$.
Notice the difference between $\omega$ and $\kappa$, the later 
appearing in the finite temperature model, for which,
there is no restriction on the values that $\gamma$ can take,
while in this model, $\gamma$ is restricted to values satisfying the inequality:
$\gamma^2 = \beta^2/4  <  1$.  The dynamical map which moves a
certain point described by the vector $\vec{r}(\tau) = (k(\tau),s(\tau))$ at the
time $\tau$  to the point $\vec{r}(\tau+\sigma) = (k(\tau + \sigma),s(\tau+\sigma))$
at time $\tau + \sigma$ is described by  
$\vec{r}(\tau+\sigma)=\bM(\sigma)\vec{r}(\tau)$
where  the matrix $\bM(\sigma)$ has the following form:
\begin{equation}\label{mmatap}
\bM(\sigma)=\rme^{{\beta\over 2}{ \sigma}}\left( \begin{array}{cc}
\cos\omega\sigma-{\beta\over 2 \omega}\sin\omega\sigma & {1\over \omega}\sin\omega\sigma \\
{-1\over \omega}\sin\omega\sigma & \cos\omega\sigma+{\beta\over 2 \omega }\sin\omega\sigma 
\end{array}\right)
\end{equation}
As for the finite temperature case, the map described by $\bM$ is also reversible,
\ie~$\bM^{-1}(\tau) = \bM(-\tau)$
and also posses the group properties;: $\bM(\tau)\bM(\tau') = \bM(\tau + \tau')$.
Integration of the third equation (\ref{ecparcl3}) 
is done from an initial time  $\tau$ to a final time 
$\tau= \tau + \sigma$ as:
\begin{equation}\label{int3cl}
\int_{\rmw(\tau)}^{\rmw(\tau+\sigma)}{\rmd\rmw\over \rmw}=-D\beta\int_{\tau}^{\tau + \sigma }\rmd\tau '\, s^2(\tau ')\,,   
\end{equation}
and with the help of the map described by the matrix $\bM$,
it is possible to write down  
$s(\tau ') = M_{21}(\tau ' - \tau) k(\tau) + M_{22}(\tau' - \tau) s(\tau)$ 
which by substituting this expression into the right hand of (\ref{int3cl}) and doing a suitable change of variable, 
one can write the solution for the chord function in this model as:
\begin{equation}\label{solmeap}
\rmw(\vec{r},\tau + \sigma) = 
\rmw \left(\, \bM^{-1}(\sigma)\; \vec{r},  \tau\,\right)
\rme^{-D\beta\left[\, \vec{r}^{T} \, \bA(\sigma)\, \vec{r}\, \right]}
\end{equation}
where $\bA(\sigma)$ is a symmetric two by two matrix whose elements are given by 
\begin{eqnarray}\label{a11}
A_{11}(\sigma)&=&\int_{0}^{\sigma} \rmd \sigma'\, M^2 _{21}(-\sigma'),\\\label{a12}
A_{12}(\sigma)&=& \int_{0}^{\sigma} \rmd \sigma'\,  M _{21} (-\sigma')M_{22}(-\sigma'),\\\label{a22}
A_{22}(\sigma)&=& \int_{0}^{\sigma} \rmd \sigma'\,  M_{22}^2(-\sigma').
\end{eqnarray}
This solution represents 
the evolution of chord function from a certain initial time $\tau$
to a final time $\tau + \sigma$.
The Wigner function representation of the solution  when an initial coherent state 
is assumed is given by:
\begin{eqnarray}\nonumber
 \mathrm{W}(q,p,\sigma) &=& {1\over \sqrt{4 \xi_1(\sigma)\xi_3(\sigma) - \xi^2_2(\sigma)}}
 \exp\left[ -  {  \xi_1(\sigma)(p-p_o(\sigma))^2  + \xi_2(\sigma)(p-p_o(\sigma))(q-x_o(\sigma))+\xi_3(\sigma)(q-x_o(\sigma))^2 
 \over 4\xi_1(\sigma)\xi_3(\sigma) - \xi^2_2(\sigma)}\right]\\\label{Wcl}
\end{eqnarray}
where $x_o(\sigma)$ and $p_o(\sigma)$ are the components of the vector $\vec{r}_o=(x_o,p_o)$
that has been mapped with the transpose of the matrix $\bM(-\sigma)$, 
this is $\vec{r}_{o}(\sigma) = \bM^{T}(-\sigma)\vec{r}_o$, and the functions $\xi_i(\sigma)$ ($i=1,2,3$), are 
given by:
\begin{eqnarray}
 \xi_1(\sigma) &=& D\beta A_{11}(\sigma) + {1\over 4}\left(M^2_{11}(-\sigma) + M^{2}_{21}(-\sigma)\right)\,,\\
 \xi_2(\sigma) &=& 2 D\beta A_{12}(\sigma) +{1\over 2}\left( M_{11}(-\sigma)M_{12}(-\sigma) + M_{21}(-\sigma)M_{22}(-\sigma)\right)\,,\\
 \xi_3(\sigma) &=& D\beta A_{22}(\sigma) + {1\over 4}\left(M^2_{12}(-\sigma) + M^{2}_{22}(-\sigma)\right)\,.
\end{eqnarray}
The time evolution of the energy can be obtained by using equation (\ref{aven}), yielding:
\begin{eqnarray}\label{ecl}
 E (\sigma) &=& E_{\mt o}\, \rme^{-\beta\sigma}\,
 \left(\cos^2\omega \sigma + {1+\beta^2/4\over 1-\beta^2/4}\sin^2\omega\sigma\right) + D\beta\left(A_{11}(\sigma) + A_{22}(\sigma)\right)\,, 
 \end{eqnarray}
where $E_{\mt o}$ is the energy of the initial condition of the oscillator 
and $A_{11}(\sigma)$ and 
$A_{22}(\sigma)$ are matrix elements of the matrix $\bA(\sigma)$, given at 
(\ref{a11}) and (\ref{a22}) respectively. 
In the long time limit the matrix $\bA$ has the following limit: 
$\lim_{\sigma\rightarrow \infty} \bA(\sigma) \rightarrow \One/(2\beta)$,
with $\One$ being the two by two unity matrix. The stationary 
state is therefore reached in this limit having the following form:
$\rmw_{\mt{st}}(\vec{r}) = 
\exp\left[-{D\over 2} \left( k^2   +  s^2\right)\right]$,
and the finite temperature model in its high temperature 
limit and the Caldeira-Leggett  model, have the same stationary form.
\subsubsection{\label{odht}Over damped case}
\begin{figure}[!htbp]
  \begin{center}
  \resizebox{138mm}{!}{\includegraphics{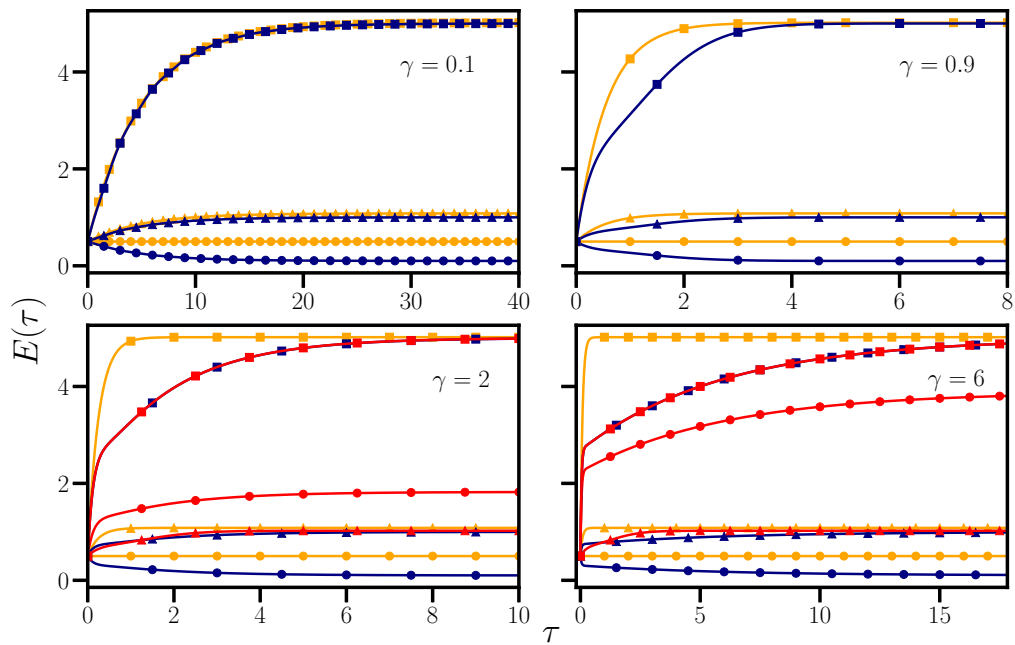}}
  \end{center}
  \caption{\label{fig1} The figures show the transient of the energy of the harmonic oscillator oscillator
  to the stationary state when initially was found in the ground state. 
  The curves depicted in blue corresponds to the 
  finite temperature model described by equation (\ref{eft});
  the curves depicted in yellow corresponds to the under damped Caldeira-Leggett model described by equation (\ref{ecl})
  and the curves depicted in red corresponds to the over damped Caldeira-Leggett model described by (\ref{eclod}).
  The curves marked with circles corresponds to a temperature of the environment of $D=0.1$ which 
  corresponds to an energy of the environment below the ground state of the oscillator; 
  the curves marked with triangles corresponds to a temperature of the environment of $D=1$ and 
  the curves marked with squares corresponds to $D=5$. }
\end{figure}

The applicability of the Caldeira-Leggett model
relies on the fact that the relation
$\gamma \hbar < \omo \hbar << k_B T$ is always fulfilled which 
in turn, the damping rate becomes indirectly 
settled in in the under damped regime. 
A consequence of this fact 
can be observed in the term 
term concerned to the 
damping in the
Caldeira-Leggett master 
equation, $\frac{\beta}{2}\;[\h{x}, \{\h{p},\varrho\} ]$ 
which does not posses a Lindblad form~\cite{Br02},
nevertheless, the influence of this term is  
minimal for large temperatures and small damping rates
and hence one can argue no violation off 
the positivity. Furthermore, it is possible to bring 
the Caldeira-Leggett master equation into 
a Lindblad form by including an additional term 
by hand into the dissipator of the Caldeira-Leggett master equation.
This term is called 
the minimally invasive term~\cite{Fe09}, and depends
inversely  on the temperature and hence the dub of
minimally invasive term
for large temperatures.
Ankerhold~\cite{An03} has derived a Fokker-Planck equation
for the Wigner function ($\mt{W}(q,p; \tau)$), at the strong
friction limit, in which the inequality
$\gammo/\omo^2 >>\hbar/k_BT,1/\omega_c,1/\gammo$ (with $\omega_c$
is a certain cut-off frequency) is fulfilled. 
In the dimensionless description described above, and for the harmonic oscillator case,
the master equation for the Wigner function in the Ankerhold model is:
\begin{equation}\label{wignerod}
 \partial_t \mt{W} = \left\{\partial_p ( q + \beta p) - p \partial_q + \beta\Omega \partial^2_p
 + \Gamma\,\partial^2_{qp}\right\}\mt{W}
\end{equation}
where $\beta = 2\gamma$ and $\Gamma = D +\Lambda - \Omega$ and  in
the limit of high temperatures, $\Omega \sim D$ and $\Lambda \sim 1/(12 D)$, while for small 
temperatures,  $\Omega \sim \beta/\pi\,  \log(\omega_c/\beta)$ and 
$\Lambda \sim 1/\beta\pi \, \log(\beta/2\pi D)$.
The first three terms in the r.h.s of equation (\ref{wignerod})
corresponds exactly to the  standard Caldeira-Leggett model, when 
the replacement of the diffusion constant $D\rightarrow \Omega$ is done; 
the last term is the new contribution due to the strong friction. 
The transformation of equation (\ref{wignerod}) into the chord function
can be easily done by taking the inverse Fourier transform 
of equation (\ref{wignerod}) as stated in (\ref{wtr}); nevertheless, 
the only term one really needs to transform is the last term of
(\ref{wignerod}), since the remaining terms have been already transformed 
in the Caldeira-Leggett model. Therefore, by 
doing so, we arrive to the following master equation 
for the chord function:
\begin{equation}\label{clchordod}
\partial_{\tau}\rmw+(\beta s-k)\partial_s\rmw
+s\partial_k\rmw=-\Omega\beta s^2\rmw -\Gamma\, k\,s\, \rmw \,.
\end{equation}
The solution for this 
master equation can be obtained by following similar lines as
the one followed when solving the Caldeira-Leggett model. First 
we write the first order partial differential equation (\ref{clchordod})
in a parametric form; 
\begin{eqnarray}\label{ecparclo1}
 {\rmd k\over \rmd \tau}&=&s,\\\label{ecparclo2}
 {\rmd s\over \rmd \tau}&=&\beta s-k,\\\label{ecparclo3}
 {\rmd \over \rmd \tau}\rmw &=& -\left( \Omega \beta s^2  + \Gamma\, k\, s\right) \rmw\,,
\end{eqnarray}
and  by coupling the first two equations, one can write a
second order ordinary differential equation for $k$:
$ \ddot{k} -\beta \dot{k} + k = 0 $,
whose solution will be given in the over damped regime by:
\begin{eqnarray}\label{ecko}
k(\tau)=\rme^{\beta \tau \over 2}\left(  a_1 \,\rme^{\mu\tau} + a_2\,\rme^{-\mu \tau}\right),
\end{eqnarray}
and as before; the solution for 
$s$ may be obtained from the first parametric equation (\ref{ecparclo2}) as:
\begin{equation}\label{ecso}
s(\tau) = \rme^{\beta \tau \over 2}\left(  a_1 \, (\mu + \beta/2)\,\rme^{\mu\tau} - a_2\,(\mu - \beta/2)\,\rme^{-\mu \tau}\right)\,,
\end{equation}
and $a_1$ and $a_2$ are the characteristic curves which remain constant for all time and 
now for the over damped regime ($\beta^2/4 >1$), we have defined 
$\mu = \sqrt{\beta^2/4 -1}$.
Thus, the dynamical map which moves a certain point described by the vector 
$\vec{r}(\tau) = (k(\tau),s(\tau))$ at the
time $\tau$  to the point $\vec{r}(\tau + \sigma) = (k(\tau + \sigma),s(\tau+\sigma))$
at time $\tau + \sigma$ is given by the following map 
$\vec{r}(\tau+\sigma)=\bN(\sigma)\vec{r}(\tau)$
where  $\bN(\sigma)$ has the following form:
\begin{equation}\label{mmatapod}
\bN(\sigma)=\rme^{{\beta\over 2}{ \sigma}}\left( \begin{array}{cc}
\cosh\mu\sigma-{\beta\over 2 \mu}\sinh\mu\sigma & {1\over \mu}\sinh \mu\sigma \\
{-1\over \mu}\sinh\mu\sigma & \cosh\mu\sigma+{\beta\over 2 \mu}\sinh\mu\sigma 
\end{array}\right)
\end{equation}
The map described by $\bN$ is also reversible, \ie $\bN^{-1}(\tau) = \bN(-\tau)$
and also posses the group properties;: $\bN(\tau)\bN(\tau') = \bN(\tau + \tau')$.
Integration of the third parametric equation (\ref{ecparclo3}) 
is done from an initial time  $\tau$ to a final time 
$\tau= \tau + \sigma$, yielding the following:
\begin{equation}\label{int3clo}
\int_{\rmw(\tau)}^{\rmw(\tau+\sigma)}{\rmd\rmw\over \rmw}
=-\int_{\tau}^{\tau + \sigma }\rmd\tau '\, \left(\Omega\beta s^2(\tau ') + \Gamma\,k(\tau')\,s(\tau')\right)\,,   
\end{equation}
and by using the map described by the matrix $\bN$,
it is possible to write down  
$k(\tau ') = N_{11}(\tau ' - \tau) k(\tau) + N_{12}(\tau' - \tau) s(\tau)$ and
$s(\tau ') = N_{21}(\tau ' - \tau) k(\tau) + N_{22}(\tau' - \tau) s(\tau)$, 
which by substituting this expression into the right hand of (\ref{int3cl}) and doing a suitable change of variable, 
one can write the solution for the chord function in this model as:
\begin{eqnarray}\nonumber
\rmw(\vec{r},\tau + \sigma) &=& 
\rmw\left(\, \bN(-\sigma)\; \vec{r}  ,\, \tau \right)
\exp\left(-\Omega\beta\, \vec{r}^{T} \, \bB(\sigma)\, \vec{r}\, -\Gamma  \vec{r}^{T} \,\bC(\sigma)\, \vec{r}\right)\\\label{solchod}
\end{eqnarray}
where $\bB(\sigma)$ and $\bC(\sigma)$ are two by two symmetric matrices whose elements are given by 
\begin{eqnarray}\label{ao11}
B_{11}(\sigma)&=&\int_{0}^{\sigma} \rmd \sigma'\, N^2 _{21}(-\sigma'),\\\label{ao12}
B_{12}(\sigma)&=& \int_{0}^{\sigma} \rmd \sigma'\,  N _{21} (-\sigma')N_{22}(-\sigma'),\\\label{ao22}
B_{22}(\sigma)&=& \int_{0}^{\sigma} \rmd \sigma'\,  N_{22}^2(-\sigma')\,,
\end{eqnarray}
and
\begin{eqnarray}\label{bo11}
C_{11}(\sigma)&=&\int_{0}^{\sigma} \rmd \sigma'\, N_{11}(-\sigma')N_{21}(-\sigma'),\\\label{bo12}
C_{12}(\sigma)&=& \int_{0}^{\sigma} \rmd \sigma'\,  \left( N _{11} (-\sigma')N_{22}(-\sigma') + N _{12} (-\sigma')N_{21}(-\sigma')\right),\\\label{bo22}
C_{22}(\sigma)&=& \int_{0}^{\sigma} \rmd \sigma'\,  N _{12} (-\sigma')N_{22}(-\sigma')\, .
\end{eqnarray}
The Wigner function representation for an initial coherent state 
is in form identical  to that of the under damped case, except that 
the components of the vector $\vec{r}_o=(x_o,p_o)$ have been mapped with the
transpose of the matrix $\bN(-\sigma)$; 
this is $\vec{r}_{o}(\sigma) = \bN^{T}(-\sigma)\vec{r}_o$ and also
the time dependent functions have now the following form:
\begin{eqnarray}
 \xi_1(\sigma) &=& \Omega\beta B_{11}(\sigma) + \Gamma C_{11}(\sigma) + {1\over 2}\left(N^2_{11}(-\sigma) + N^{2}_{21}(-\sigma)\right)\,,\\
 \xi_2(\sigma) &=& 2 \Omega \beta B_{12}(\sigma) + 2\Gamma C_{12}(\sigma) 
 +N_{11}(-\sigma)N_{12}(-\sigma) + N_{21}(-\sigma)N_{22}(-\sigma)\,,\\
 \xi_3(\sigma) &=& \Omega\beta B_{22}(\sigma) + \Gamma C_{22}(\sigma) +  {1\over 2}\left(N^2_{12}(-\sigma) + N^{2}_{22}(-\sigma)\right)\,.
\end{eqnarray}
The average energy can also be calculated with the help of equation (\ref{aven}) yielding for this case:
\begin{eqnarray}\nonumber
 E (\sigma) &=& E_{\mt o}\, \rme^{-\beta\sigma}\,
 \left(\cosh^2\mu\sigma - {1+\beta^2/4\over 1-\beta^2/4}\sinh^2\mu\sigma\right)
 + \Omega\beta\left(B_{11}(\sigma) + B_{22}(\sigma)\right) + \Gamma\, \left( C_{11}(\sigma) + C_{22}(\sigma)\right)\,.\\\label{eclod}
 \end{eqnarray}
 The only circumstantial difference in the form of the solutions 
 of the over damped and the under damped
 cases of the Caldeira-Leggett model for the high temperature regime 
 relies basically on the inclusion of the new quadratic 
 term in the exponential of (\ref{solchod}) containing the matrix $\bC$ 
 and of course the evident transition of the matrix $\bM$ to $\bN$ 
 which can be done when the angular frequency $\omega$ is taken complex 
 \ie $\omega=\rmi\mu$ for values of $\gamma$ larger than one.
 In order to see the differences between the different models,
 we have depicted the transient of the energies 
 to the stationary state in figure \ref{fig1} for the under damped and 
 over damped regimes, when the oscillator is initially in the ground state.  
 For the over damped regime, in addition to equation (\ref{eclod}), 
 we have also plotted the energy of the under damped 
 Caldeira-Leggett model when the replacement of the matrix $\bM$ to $\bN$,
 and the angular frequencies $\omega$ to $\rmi \mu$ is done.
 The energies have been plotted for different values of $D$ and damping $\gamma$ ($\beta = 2\gamma$).
 The lowest temperature we have used corresponds to an average energy of the bath which is 
 below the ground state energy of the oscillator.
 He have use this temperature to show that the under damped Caldeira-Leggett model will thermalize the 
 oscillator to a forbidden energy. Of course this problem is solved whenever 
 $D > 1/2$ or   $k_B T  >  \hbar \omega_o /2$ . 
 The finite temperature model does not has this problem as
 it keeps the oscillator at the ground state energy as one would expect.
 As the temperatures rises up,  
 the Caldeira-Leggett models and the finite temperature model
 converges to the same stationary value but through different paths. 
 When the damping is increased, the finite temperature model
 rapidly reaches the stationary value without showing 
 any dependence on the oscillator dynamics, similar to what 
 happens in the depolarizing channel model~\cite{NiCh10}; this 
 fact contrast to the Caldeira-Leggett model where
 the dependence on the oscillator dynamics at the transients is important.
 In the strong friction limit ($\gamma >1 $) the over damped Caldeira-Leggett model disagree with 
 the finite temperature and the under damped Caldeira-Leggett models 
 and converge very well for high temperatures. We believe that for the low temperature 
 rates, this disagreement is related to the strong quantum fluctuations dependence 
 argued by Ankerhold~\cite{An03}.
 We also notice that the over damped model does not seems to provide any significant
 advantage to the traditional Caldeira-Leggett model at the strong friction limit and 
 at high temperatures, as the simplest replacement of the angular frequency with the
 complex angular frequency in the under damped Caldeira-Leggett model seems to bring almost the same 
 results to those of the over damped model. 
 
\section{\label{df}Driven oscillator}
The method described above for solving the dynamics of an open quantum system 
is efficient when applied to a more complex type of systems as long as the 
dynamics associated to the Hamiltonian can be described through a dynamical map. 
In this section  we  give analytic solutions to the driven open quantum 
harmonic oscillator as an exemplification of the application of the method. 
The driving in this case is 
considered as an
external time dependent periodic force exerted on the oscillator.
The Hamiltonian of the driven oscillator has the following 
form~\cite{Pi06}: 
$ H(\tau) = H_{\mt {osc}}(\tau)  - \lambda(\tau)\,\hat{x},$ 
where 
\begin{equation}\label{lt}
\lambda(\tau) = \lambda \cos\nu \tau \,,
\end{equation}
thus, the master equation we will solve is: 
$\rmi\, \rmd\varrho/\rmd \tau =
[H_{\mt{osc}} - \lambda(\tau)\,\hat{x},\varrho] + \rmi \mathcal{A}[\varrho]$,
where the dissipator $\mathcal{A}[\varrho]$ will be either 
for the high or finite temperature models.
Following similar lines as for the previous cases
we can derive an analytic solution for the master equation
of the driven oscillator, when coupled to either 
a finite temperature or a high temperature bath.
Independently of which type of bath one assumes,
the only differences to the previous master equations relies on the 
driving term of the Hamiltonian of the oscillator,
which when transformed into the chord function 
representation this term simply becomes: 
$\lambda(\tau) \,s \,\rmw(k,s,\tau)$. 
This additional term must
be included in the third parametric equations (\ref{paraft3}) or (\ref{ecparcl3}),
as it does not posses any partial derivative with respect to $k$
or $s$. In the following we present the analytic solutions for these
two models of dissipation.
\subsubsection{Finite temperature case:}
In the chord function representation, the master equation for the finite temperature 
case is written as:
\begin{eqnarray}\nonumber
 \partial_{\tau}\rmw  + (s + \gamma k )\partial_k \rmw - (k-\gamma s)\partial_s\rmw 
 &=& -\left( \rmi \lambda(\tau)\,s  + {\gamma_{+}\over 2} (k^2 + s^2)\right)\rmw\,.\\\label{mechftd} 
\end{eqnarray}
The solution for this non-autonomous partial differential equation 
can be derived following the same lines as done in the previous cases, yielding:
\begin{eqnarray}\nonumber
\rmw(\vec{r},\tau+\sigma)&=& 
\rmw\big(\, \bR(-\sigma) \vec{r}\, ,\, \tau \big)
\exp\left( - {\gamma_{+}\over 2}\,\alpha(\sigma)\, |\vec{r}\,|^2  - \rmi\, \vec{\eta}(\tau + \sigma)\cdot \vec{r}   \,  \right)
\\\label{solwftd}
\end{eqnarray}
where we have defined the vector $\vec{\eta}(\tau + \sigma) = (\eta_1(\tau+\sigma) , \eta_2(\tau + \sigma))$
whose components are given by:
\begin{eqnarray}\label{eta1}
 \eta_1(\tau + \sigma) &=&\int_{0}^{\sigma} \rmd \sigma'\,\lambda(\tau+\sigma - \sigma') R_{21}(  -\sigma')\\\nonumber
  & = & {\lambda\over 2} \rme^{-\gamma \sigma}\left\{  {1\over |\Delta_-|^2}
  \mathrm{Re}\left[\Delta_-\rme^{-\rmi\nu(\tau+\sigma)+\rmi(\nu-1)\sigma}\right]
    + {1\over |\Delta_+|^2}\mathrm{Re}\left[\Delta_+\rme^{-\rmi\nu(\tau+\sigma)+\rmi(\nu+1)\sigma}\right]  \right\}\\\nonumber
    &&-{\lambda\over |\Delta_+\Delta_-|^2}\mathrm{Re}\left[\Delta_+\Delta_-\rme^{-\rmi\nu(\tau+\sigma)}\right]
    \end{eqnarray}
    \begin{eqnarray}\label{eta2}
 \eta_2(\tau + \sigma) &=&\int_{0}^{\sigma} \rmd \sigma'\,\lambda(\tau+\sigma - \sigma') R_{22}(  -\sigma')\\\nonumber
 & = & {\lambda\over 2} \rme^{-\gamma \sigma}\left\{  {1\over |\Delta_-|^2}
 \mathrm{Im}\left[\Delta_-\rme^{-\rmi\nu(\tau+\sigma)+\rmi(\nu-1)\sigma}\right] 
   + {1\over |\Delta_+|^2}\mathrm{Im}\left[\Delta_+\rme^{-\rmi\nu(\tau+\sigma)+\rmi(\nu+1)\sigma}\right]  \right\}\\\nonumber
    &&-{\lambda\over |\Delta_+\Delta_-|^2}\mathrm{Im}\left[\left(\nu + \rmi \gamma\right)\Delta_+\Delta_-\rme^{-\rmi\nu(\tau+\sigma)}\right]
\end{eqnarray}
whit $\Delta_+ = \nu+1 - \rmi\gamma$ and $\Delta_- = \nu-1 - \rmi\gamma$ and $R_{21}$ and $R_{22}$
are components of the map $\bR$ given at (\ref{Matft}). Notice that the contribution 
of the driving in the chord function solution 
appears as an extra time dependent phase additional to the solution previously obtained for this 
type of model. The Wigner function can be easily obtained by doing the inverse Fourier transform of 
the solution and becomes particularly simple to obtain if the initial state 
is considered as before a general coherent state. In fact, the form of the Wigner function is 
the same as the one described in (\ref{Wft}) but now $x_o(\sigma)$ and 
$p_o(\sigma)$ are the components of the vector $\vec{r}_o=(x_o,p_o)$
that has been transformed as: $\vec{r}_{o}(\sigma) = \bR^{T}(-\sigma)\vec{r}_o  - \vec{\eta}(\tau + \sigma) $.
The average energy of the oscillator can be calculated
by employing equation (\ref{aven}), yielding
\begin{eqnarray}\label{eftd}
 E (\sigma)& =& E_{\mt o}\, \rme^{-2\gamma\sigma} + \left(\bar{n} + 1/2\right)\left(1- \rme^{-2\gamma \sigma}\right)
 + {1\over 2}\left(\eta^2_1(\tau+\sigma) + \eta^2_2(\tau+\sigma)  \right)\,, 
 \end{eqnarray}
which is basically the energy of the finite temperature model plus the contribution 
 due to the driving force.
 
\subsubsection{The Caldeira-Leggett case:}
\begin{figure}[!htbp]
  \begin{center}
  \resizebox{140mm}{!}{\includegraphics{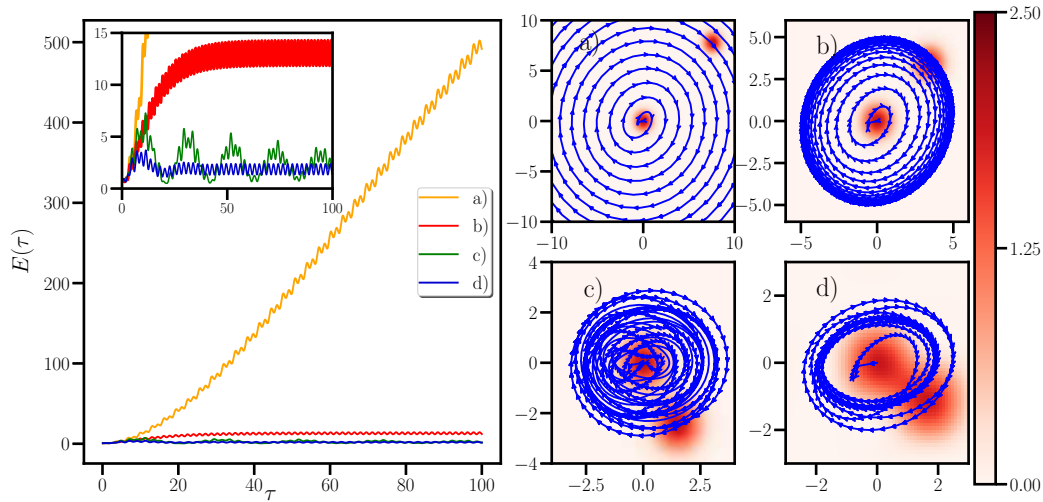}}
  \end{center}
  \caption{\label{fig2} The figure shows the behavior of the energy and the path of the system in the 
  Wigner function representation for the resonant case: a) $\nu=1$ $\gamma=0.01$ and b) $\nu=1$ with $\gamma = 0.1$
  and the cases out of resonance: c) $\nu = 0.7$ and $\gamma = 0.01$ and d) $\nu=0.7$ and $\gamma = 0.1$ for the finite temperature model 
  at the zero temperature limit. 
  In the figure depicted at the right, the system, initially at the ground state is represented by the coherent state 
  located at the origin. The driving moves the coherent state keeping
  it coherent along the trajectories represented by the blue arrows. In these figures we have depicted the location of the coherent 
  state after some time of evolution when it has move along the trajectory.}
\end{figure}

\begin{figure}[!htbp]
  \begin{center}
  \resizebox{140mm}{!}{\includegraphics{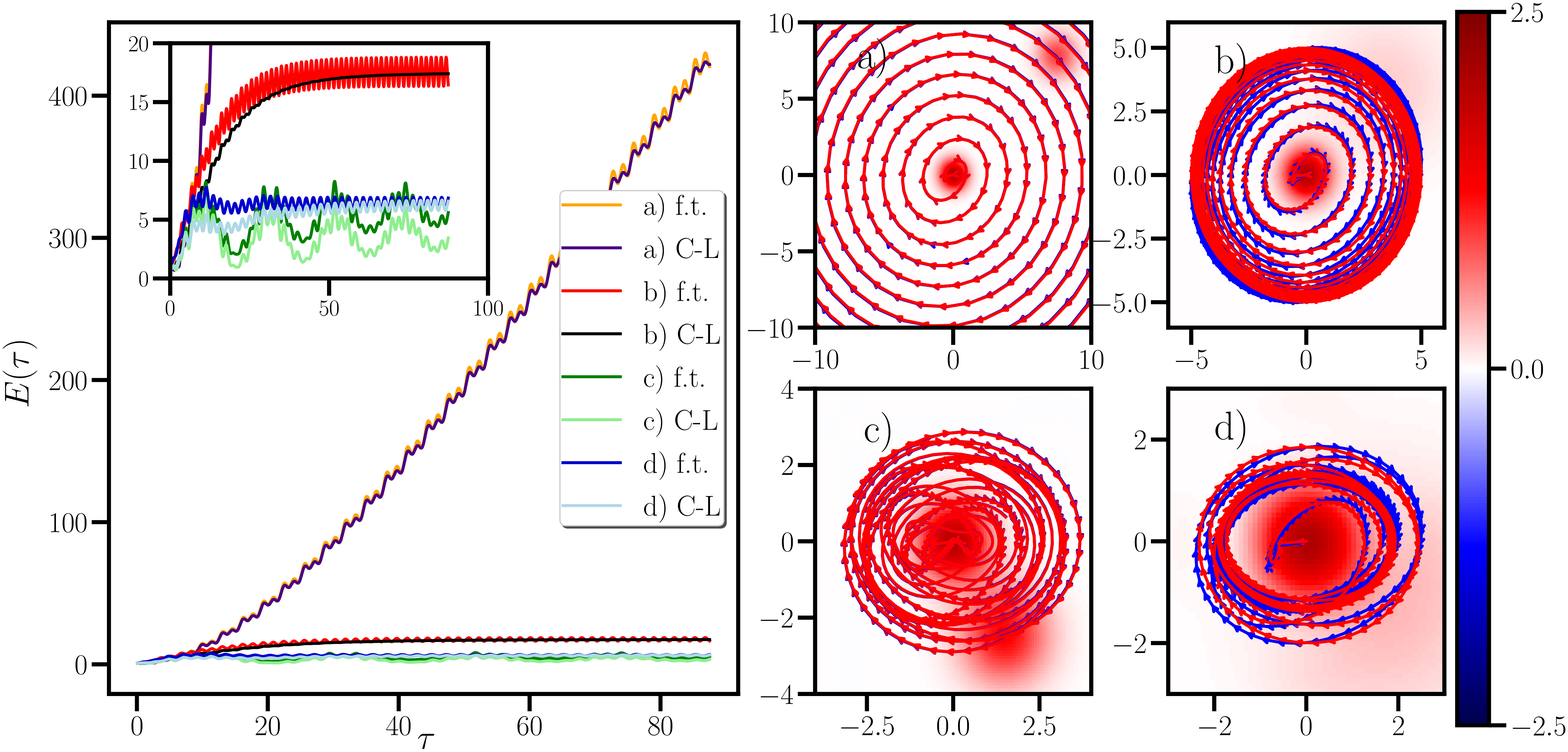}}
  \end{center}
  \caption{\label{fig3}The figure shows the behavior of the energy and the path of the system in the 
  Wigner function representation for the resonant case: a) $\nu=1$ $\gamma=0.01$ and b) $\nu=1$ with $\gamma = 0.1$
  and the cases out of resonance: c) $\nu = 0.7$ and $\gamma = 0.01$ and d) $\nu=0.7$ and $\gamma = 0.1$ for the finite 
  temperature (f.t.) and the Caldeira-Leggett (C-L) models and the temperature used in the figures was settled to $D=5$. 
  In the figure depicted at the right, the path that follows the initial coherent state located at the origin (the ground state)
  are represented by the blue and the red curves, the former representing the finite temperature model and the later,
  the Caldeira-Leggett model.  As the coherent states evolve in time it widens due to the influence of the temperature
  until it reaches a maximum width of the value of the temperature of the bath.}
\end{figure}

 In the chord function representation, the master equation for the finite temperature 
case may be written as:
\begin{equation}\label{mechftd}
 \partial_{\tau}\rmw  + (\beta s - k)\partial_s\rmw +s \partial_k \rmw  
 = -\left( \rmi \lambda(\tau)\,s + D\,\beta\, s^2\right)\rmw \,. 
\end{equation}
The solution again can be obtained following the same steps as before yielding the following
for the chord function:
\begin{eqnarray}\nonumber
\rmw(\vec{r},\tau + \sigma) &=& 
\rmw\left( \bM(-\sigma) \vec{r}  ,\, \tau \right)
\exp\left( -D\beta \, \vec{r}^{T} \, \bA(\sigma)\, \vec{r}\, - \rmi\, \vec{\xi}(\tau + \sigma)\cdot \vec{r}\right)\\\label{solchd}
\end{eqnarray}
where we have defined the vector $\vec{\xi}(\tau + \sigma) = (\xi_1(\tau+\sigma) , \xi_2(\tau + \sigma))$
whose components are obtained through:
\begin{eqnarray}\label{xi1}
 \xi_1(\tau + \sigma) &=&\int_{0}^{\sigma} \rmd \sigma'\,\lambda(\tau+\sigma - \sigma') M_{21}(  -\sigma')\\\nonumber
  & = & {\lambda\over 2 \omega } \rme^{-{\beta\over 2 } \sigma}\left\{  {1\over |\Delta'_-|^2}
  \mathrm{Re}\left[\Delta'_-\rme^{-\rmi\nu(\tau+\sigma)+\rmi(\nu-\omega)\sigma}\right]
    + {1\over |\Delta'_+|^2}\mathrm{Re}\left[\Delta'_+\rme^{-\rmi\nu(\tau+\sigma)+\rmi(\nu+\omega)\sigma}\right]  \right\}\\\nonumber
    &&-{\lambda\, \omega \over |\Delta'_+\Delta'_-|^2}\mathrm{Re}\left[\Delta'_+\Delta'_-\rme^{-\rmi\nu(\tau+\sigma)}\right]
  \end{eqnarray}  
  \begin{eqnarray}\label{xi2}  
 \xi_2(\tau + \sigma) &=&\int_{0}^{\sigma} \rmd \sigma'\,\lambda(\tau+\sigma - \sigma') M_{22}(  -\sigma')\\\nonumber
 & = & -{\beta\over 2}\xi_1(\tau+\sigma)\\\nonumber
  &&+ {\lambda \over 2\omega} \rme^{-{\beta\over 2} \sigma}\left\{  {1\over |\Delta'_-|^2}
 \mathrm{Im}\left[\Delta'_-\rme^{-\rmi\nu(\tau+\sigma)+\rmi(\nu-\omega)\sigma}\right] 
 + {1\over |\Delta'_+|^2}\mathrm{Im}\left[\Delta'_+\rme^{-\rmi\nu(\tau+\sigma)+\rmi(\nu+\omega)\sigma}\right]  \right\}\\\nonumber
    &&-{\lambda\over |\Delta'_+\Delta'_-|^2}\mathrm{Im}\left[\left(\nu + \rmi \gamma\right)\Delta'_+\Delta'_-\rme^{-\rmi\nu(\tau+\sigma)}\right]
\end{eqnarray}
where now $\Delta'_+ = \nu+\omega - \rmi\gamma$ and $\Delta'_- = \nu-\omega - \rmi\gamma$ and 
$M_{21}$ and $M_{22}$ are the components of the map described in (\ref{mmatap}). 
Similar to the  finite temperature model, the effect of the driving is noticed in 
an extra time dependent phase in the chord function and also the Wigner function 
of the system can be obtained by doing the inverse Fourier transform that will yield 
a solution with the same form to that given at (\ref{Wcl}), but now 
$x_o(\sigma)$ and $p_o(\sigma)$ are the components of the vector $\vec{r}_o=(x_o,p_o)$
that has been transformed according to: 
$\vec{r}_{o}(\sigma) = \bR^{T}(-\sigma)\vec{r}_o - \vec{\xi}(\tau + \sigma)$.
The average energy of the oscillator in this case has the following form:
\begin{eqnarray}\nonumber
 E (\sigma) &=& E_{\mt o}\, \rme^{-2\beta\sigma}\,
 \left(\cos^2\omega \sigma + {1+\beta^2/4\over 1-\beta^2/4}\sin^2\omega\sigma\right) + D\beta\left(A_{11}(\sigma) +
 A_{22}(\sigma)\right) + {1\over 2}\left(\xi^2_1(\tau+\sigma) + \xi^2_2(\tau+\sigma)  \right)\,. \\\label{ecld}
 \end{eqnarray}
In both cases, the finite temperature and the Caldeira-Leggett model, the 
contribution of the driving is only represented 
in global time dependent phases which multiply the chord function 
of the open system previously obtained. Particularly for an initial coherent state,
the effect is to drive the system along trajectories described 
by $(x_o(\sigma) - \eta_1(\tau + \sigma), p_o(\sigma) - \eta_2(\tau + \sigma) )$ for the finite temperature model 
and $(x_o(\sigma)-\xi_1(\tau + \sigma), p_o(\sigma)-\xi_2(\tau + \sigma) )$ for the Caldeira-Leggett model.
On the other hand, this implies  that the 
effects of the bath and the driving seems to be decoupled in the sense that 
the action of the bath is always directed to damp and thermalize the system 
while the influence of the driving is to
drive the system along these trajectories. 
In figure \ref{fig2} we have depicted the zero temperature
limit of the finite temperature 
model for an initial ground state of the oscillator. We show 
the energy of the oscillator and the trajectory a coherent state will follow,
for two values of the frequency of the driving, 
one of them being the resonant frequency, and two values of damping. 
In the figure, one can observe that at small damping rates and at the resonant frequency, (case a),
the energy increases very rapidly and the Wigner function is driven out far from the ground state (the origin)
but non of the energy provided by the driving is used to widen the Gaussian state \ie, the initial state 
remains coherent at all time. This is at a certain point what one would expect when the 
range of energies needed to generate the diffusion of the wave packet are much more 
smaller to the energy provided by the driving. In the case b) the frequency of the driving is also kept 
at the resonant frequency but the damping has been increased. In this situation the system reaches a quasi stationary 
condition with an averaged stationary energy and a quasi stationary orbit of the coherent state. For case
c) we have moved the frequency of the driving below the resonant frequency and used a small damping. Here we observed 
that the coherent state is sometimes moved away from the ground state and sometimes is driven to the origin again, 
since the driving is out of resonance and thus it could slow down or speed up the oscillator.
When the damping is increased as seen at case d) then the orbits become more regular as the damping 
begin to dominate the driving.  In figure \ref{fig3}, we have sketched the same type
of figure as figure \ref{fig2} but now for a temperature different to zero and
including the Caldeira-Leggett model as well. 
In this case, the effect of the temperature in the Wigner function is to
widen the initial coherent state
to a Gaussian state with width equal to the temperature, 
corresponding this state to a mixed thermal state in the energy basis,
while the driving displace the Gaussian out of 
the origin along the same paths as before.  For small rates of damping, 
the driving dominates and the trajectory of the coherent state is basically the same 
for both models of dissipation. Also at this regime of low damping, the coherent state 
last more time localized, and for higher rates of damping, the trajectories start to 
present differences and the system becomes more rapidly delocalized.

\section{\label{sum}Summary}
We have used the Fourier transform of the Wigner function or chord function, 
to solve analytically the quantum open harmonic oscillator 
for the most used models of dissipation, such as the zero-finite
temperature model, based on the optical master equations possessing a Lindblad 
form, and the high temperature model for the under damped and over damped 
regimes, based on the Caldeira-Leggett master equations. We have derived the 
the solutions by first obtaining the dynamical map associated to the 
master equation in the chord function representation, which has the form 
of an evolution matrix of a dissipative classical system, possessing 
group properties. 
We have presented a comparison 
of the different models in the behavior of the transients of the 
energy to the stationary state. We  have shown the regime 
at which the under damped Caldeira-Leggett model seems to fail 
and stated the most circumstantial differences among the models.
Additionally we found for the over damped 
Caldeira-Leggett model does not reproduces the thermalization dynamics for 
low temperatures as the finite temperature model does;  something that 
may be related to the strong quantum fluctuations dependences as stated by Ankerhold.
At high temperatures, the under damped and over damped models have very
similar behaviors and in principal, we do not see any real advantage 
or strong justification on why to use the Overdamped model over the underdamped model 
when the replacement of the characteristic frequency of the dissipative 
oscillator is replaced by its complex description \ie $\omega \rightarrow \rmi \nu$.
On the other hand the finite temperature model gives
correct results at low temperatures and high temperatures in the sense that 
it reaches what one  would think to be the  stationary state, nevertheless
the transients of the system to the stationary state for this model 
does not seems to have any dependence on the oscillator dynamics 
and for large damping, it behaves as a depolarizing channel model of dissipation. 
We have also derived and studied the solutions of the 
driven open harmonic oscillator model when 
the driving of the oscillator corresponds to the application of 
an external time dependent and periodic force. 
We have shown 
that the action of the dissipative mechanism is always to get rid of 
the dependence on the  initial conditions, the decoherence, and to 
lead the system into a thermal state configuration;
while the driving moves the state of the system to different 
places of the energy configuration states depending on the 
driving force parameters and the damping rates. The system 
reaches quasi stationary states in the long time limit 
where the final configuration depends mainly on
the frequency of the driving force and the damping rates.

\section{Acknowledgments}
PCLV thanks to SNI-CONACyT for the support and  RSS acknowledges the
support of the  NPTC-PRODEP project UDG-PTC-1368 and UDG-CA-959.

\bibliographystyle{apsrev4-1}
\bibliography{man}

\end{document}